# Graphenic Carbon-Silicon Contacts for Reliability Improvement of Metal-Silicon Junctions

Max Stelzer[1], and Franz Kreupl[1]
[1]Technical University of Munich, Arcisstr. 21, 80333 Munich, Germany, email: max.stelzer@tum.de

***Abstract* –** Contact resistance and thermal degradation of metal-silicon contacts are challenges in nanoscale CMOS as well as in power device applications. Titanium silicide (TiSi) contacts are commonly used metal-silicon contacts, but are known to diffuse into the active region under high current stress. In this paper we show that a graphenic carbon (C) contact deposited on n-type silicon (C-Si) by CVD, has **the same low Schottky barrier height of 0.45 eV as TiSi**, **but a much improved reliability against high current stress**. **The C-Si contact is over $10^8$ times more stable against high current stress pulses than the conventionally used TiSi junction. The C-Si contact properties even show promise to establish an ultra-low, high temperature stable contact resistance.** The finding has important consequences for the enhancement of reliability in power devices as well as in Schottky-diodes and electrical contacts to silicon in general.

## I. Introduction

Metal to semiconductor contacts are essential elements in integrated and discrete electronic devices. Current state-of-the-art contacts to FinFETs rely on titanium silicide (TiSi) formation [1] and existing efforts to reduce the contact resistance make heavy use of TiSi [2,3]. TiSi is also used in Schottky diodes for zero-barrier mixer applications and up to THz frequencies [4,5]. Joule heating and electric fields of sufficient magnitude can force the migration and diffusion of metal ions into the underlying semiconductor [5,6]. Especially in an electrostatic discharge (ESD) event, a high current is flowing through the device causing a degradation of the inherent electrical properties or even leading to a junction burnout and consequently a failure of the device.

A Schottky diode structure, as shown in Fig.1 is used in this work to compare the characteristics and reliability of a TiSi-based diode (commercial BAT15 diode from Infineon) with a C-Si Schottky diode, where the “metallic contact” to the silicon is established by graphenic carbon (C) deposited by CVD [7,8]. The same silicon vehicle with all dimensions, dopants and guard rings like the commercial BAT15 was provided by the manufacturer and was used in the experiment to allow for a direct comparison. Due to the small thickness of 150 nm of the $n^-$-epitaxial layer, the BAT15 vehicle is very susceptible to degradation in its electrical reverse-bias characteristics when contaminations like metal ions enter the epitaxial region, like shown in Fig. 2(a). By studying the reverse characteristic of the Schottky diode at -1V reverse bias after the application of a short, 100 ns long, high current pulse, a damaged diode can easily be identified by an increase of the reverse current. The power per volume in a conductor during a current pulse is given by $j \cdot E = j^2 \cdot \rho$, where $j$ is the current density, $E$ is the local electric field and $\rho$ the electrical resistivity. The magnitude of the applied current density has the strongest influence on the induced damage to the device and hence on its reliability.

## II. TiSi-diodes

Commercially available BAT15 Schottky diodes with TiSi contacts were pulsed with different current densities and the pulse number dependent reliability results are plotted in Fig. 3. The current pulse with duty cycle smaller than 0.0001 has a length of 100 ns and a rise/fall time of about 15 ns, like shown in Fig. 2(b). If the reverse current reaches a value of 100 µA at -1 V, the on/off ratio of the diode is deteriorated by a factor of 100 and the diode is declared as failed. From Fig. 3 it is observable that the maximum number of transient stress pulses depends on the used current density level. For 1.35 MA/cm$^2$, which is slightly above the permissible pulse load of 1.21 MA/cm$^2$, specified by the manufacturer, the diode is able to withstand up to $10^6$ events but the failure probability is spread over a wide range. Moving to higher current densities increases the probability for damage severely and **the TiSi junction can only withstand 2-4 pulses at 3.5 MA/cm$^2$.**

## III. Graphenic Carbon-Silicon Contacts

C-Si diodes were fabricated by CVD-deposition of graphenic carbon onto the cleaned Si-surface of the BAT15 structure at temperatures between 800°C and 1000°C [7,8]. The layered structure of the carbon layer is nicely seen in the cross-section shown in Fig. 4(b). The cleaved sample reveals the high anisotropy of the carbon growth. Consequently, the graphenic carbon has an in-plane resistivity of 1 mΩ·cm and in the direction perpendicular to the surface a resistivity of 50 mΩ·cm. This value is extracted from analyzing samples with varying carbon thickness and fitting to the obtained data plotted in Fig. 4(a). The graphenic carbon film is structured by depositing a metal stack consisting of 50 nm Ti, 1.3 µm Cu and 40 nm Au through a shadow mask on to the carbon. Subsequently the carbon was etched away in a hydrogen plasma with the metal stack acting as a hard mask (Fig. 1(b)).

## IV. Characteristics and Reliability of C-Si Diodes

The dc-characteristic of the pristine (no current pulse yet applied) TiSi and C-Si diodes with a carbon thickness of 58 nm are compared in Fig. 5. The substrate of the C-Si is not back-thinned like in the commercial TiSi-diode and therefore the C-Si diode has a little higher resistance. The ideality factors $n$ and the Schottky barrier heights (SBH) $\Phi_B$ of the two diodes are almost the same. Fig. 5 also displays the dc-characteristics after three current pulses for the TiSi-diode and after 100 million pulses for the C-Si diode with a pulse strength of 3.5 MA/cm$^2$. After three pulses the TiSi diode is destroyed as evidenced by the high reverse current. In contrast to this, the C-Si diode survives at least 100 million equivalent current pulses. After 100 million pulses the reverse current is reduced by a factor of 4 and the Schottky barrier height $\Phi_B$ increased. This might be due to the formation of a very thin SiC-layer at the C-Si interface.

A graphenic carbon thickness of 28 nm was used to evaluate the impact of different current pulse lengths and fixed amplitude. The results are plotted in Fig. 6. The diodes can withstand up to 548 million pulses with a width of 100 ns before they fail. The failure probability is getting higher at a lower total number of pulses by increasing the pulse length. But the C-Si diodes still can withstand up to 16,000 pulses with a current pulse length of 1 µs which is still much more than the BAT15 can endure at a pulse width of 100 ns. Fig. 7 compares the failure probability of the TiSi BAT15 and the C-Si diode and clearly illustrates that graphenic carbon can **enhance the robustness against high current events (like ESD) by a factor of over 100 million.**

The top metallization of the diode was analyzed by a scanning electron microscope (SEM) to examine how much the surrounding area of the active device region is damaged by the current pulses (Fig. 8). For a 100 ns pulse, the damage to the top metallization is hardly visible as the event is so short that heat can hardly spread away from the active area (Fig. 8(a)). The visible bump here may just arise due to the thermomechanical stress by thermal expansion at each cycle. In Fig. 8(b-c) the top metallization is heated to the point that it is even melted and caused a pile-up of the metal. The copper starts to spit out molten metal on the periphery of the diode at a pulse width of 500 ns. Fig. 9 reveals that the failed diodes are not open devices, but have been shortened, most probably by a diffusion of the dopants. Si-C diodes with 28 nm and 58 nm carbon thickness were tested at even higher current levels. For the diode with a 58 nm thick carbon film, a maximum applicable current density of 9.3 MA/cm$^2$ could be identified. Like shown in Fig. 10, the first pulse is not destructive to the device. Only a slightly higher reverse current is observable but the acceptable breakdown voltage (100 µA @ $V_r$=4 V), as specified for the BAT15, is still fulfilled. Only the second pulse is destructive to the diode and it completely fails. For the sample with the 28 nm thick carbon a maximum manageable current strength for a single pulse was identified to be 8.0 MA/cm$^2$. The reason of the deterioration could not yet be revealed. It could be redistribution and diffusion of dopants due to the induced temperature and/or diffusion of Ti through the carbon layer towards the interface.

In order to achieve ultra-low specific contact resistances $\rho_c$ the SBH $\Phi_B$ need to be decreased. This is often achieved by inserting a thin oxide layer between metal and silicon. Unfortunately this approach is not compatible with high temperatures [2,3]. The value of the saturation current density $J_0$ of a contact is a good indication for the SBH and the obtainable $\rho_c$. Fig. 11 compares current density plots of C-Si with recent efforts from Hu et al. to achieve low $\rho_c$. **Pulse-annealed C-Si contacts clearly outperform recent efforts to achieve record low $\rho_c$ and even pristine C-Si contacts are within an attractive range.**

## V. Electro-Thermal Simulations

To get a deeper insight to the heat distribution in the diode during the ESD event, coupled time-dependent electro-thermal simulation based on finite element method (FEM) were performed. Temperature dependent parameters were used for all materials. The experimentally measured current waveform and density was used in the simulations. The temperature distribution in the C-Si diode near the active region is shown in Fig. 12(a-d) for different time steps. At low current densities the diode keeps cool and close to room temperature. At high current densities a circumferential hot spot is created around the active area close to the guard ring structure which is guided by current crowding. The temperature even rises to the melting point of silicon (Fig. 12(c)) but due to the *unusual temperature dependent electrical resistivity of carbon (which has a minimum at ~900 K)*, not enough current can be provided to keep the hot spot at high temperature. The heat in the device can spread to the periphery which reduces the maximum temperature. When the same layout is used with Ti at the interface, the hot spot doesn't disappear which confirms a higher damage potential of the sample without carbon (Fig. 12(e)). This edge behavior has been verified in the real C-Si diode by stressing the diode with a thinner top metallization. Fig. 12(f) illustrates that a recrystallization of the Cu occurs at the transition region between active and guard ring area occurs, which points to a higher temperature in this edge region and confirms the simulated results.

## Acknowledgment & Funding

The authors gratefully thank Infineon Technologies AG for providing the used silicon vehicle and W. Simbürger and J. Dietl for fruitful discussions. Supported by DFG through the TUM International Graduate School of Science and Engineering (IGSSE).

## References

[1] D. James, "Leading Edge Silicon Devices", Chipworks, AVS Ultrashallow Junction Tech Group Workshop, Semicon West 2015.

[2] H. Yu et al. "MIS or MS? Source/drain contact scheme evaluation for 7nm Si CMOS technology and beyond." 16th International Workshop on Junction Technology (IWJT). IEEE, 2016.

[3] H. Yu et al., 1.5×10-9 Ω· cm² Contact Resistivity on Highly Doped Si:P Using Ge Pre-amorphization and Ti Silicidation, IEDM 2015

[4] Z. Ahmad et al., "9.74-THz Electronic Far-Infrared Detection Using Schottky Barrier Diodes in CMOS", IEDM 2014.

[5] A. Mai et al., "Reliability aspects of TiSi-Schottky Barrier Diodes in a SiGe BiCMOS technology," IEEE, 2015

[6] K. Banerjee et.al., "Characterization of contact and via failure under short duration high pulsed current stress," IEEE, 1997.

[7] F. Kreupl, "Carbon-based Materials as Key-enabler for "More than Moore"," *MRS Proceedings*, vol. 1031, Cambridge Univ. Press, 2010.

[8] S. Huebner et al. "High Performance X-Ray Transmission Windows Based on Graphenic Carbon." IEEE Transactions on Nuclear Science 62.2 (2015): 588-593.

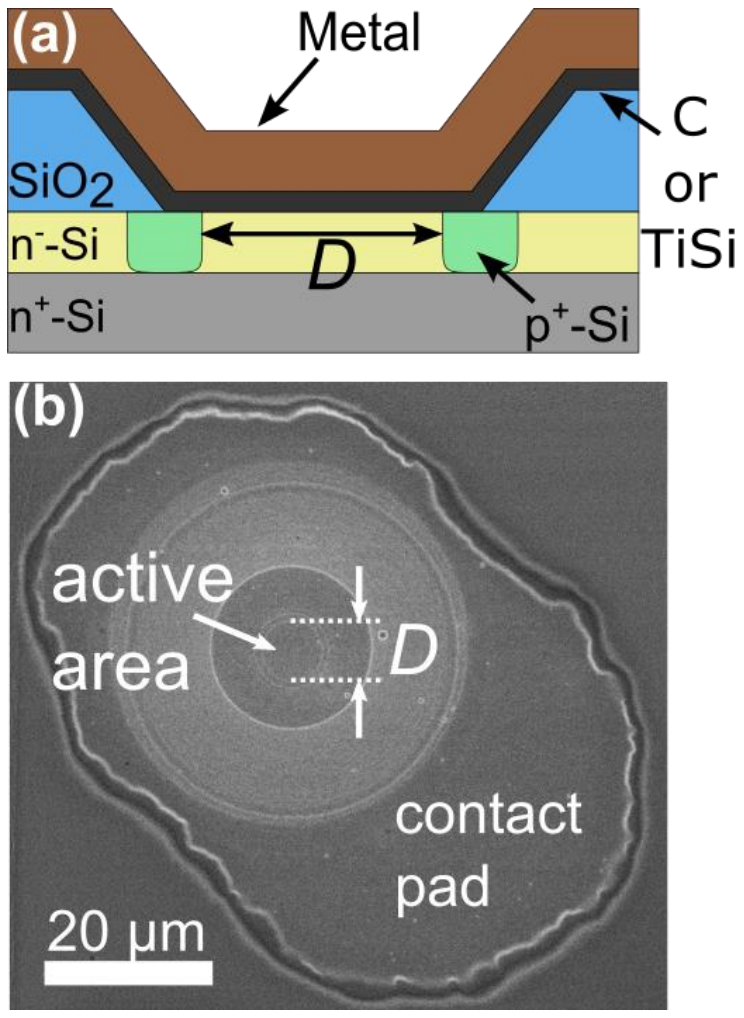

Fig. 1. (a) shows a schematic cross-section of the used Schottky diode with guard ring and C or TiSi at the interface. The SEM image in (b) shows the top view of a C-Si diode with top contact deposited through a shadow mask. The diameter $D$ of the active region is 7.6 µm.

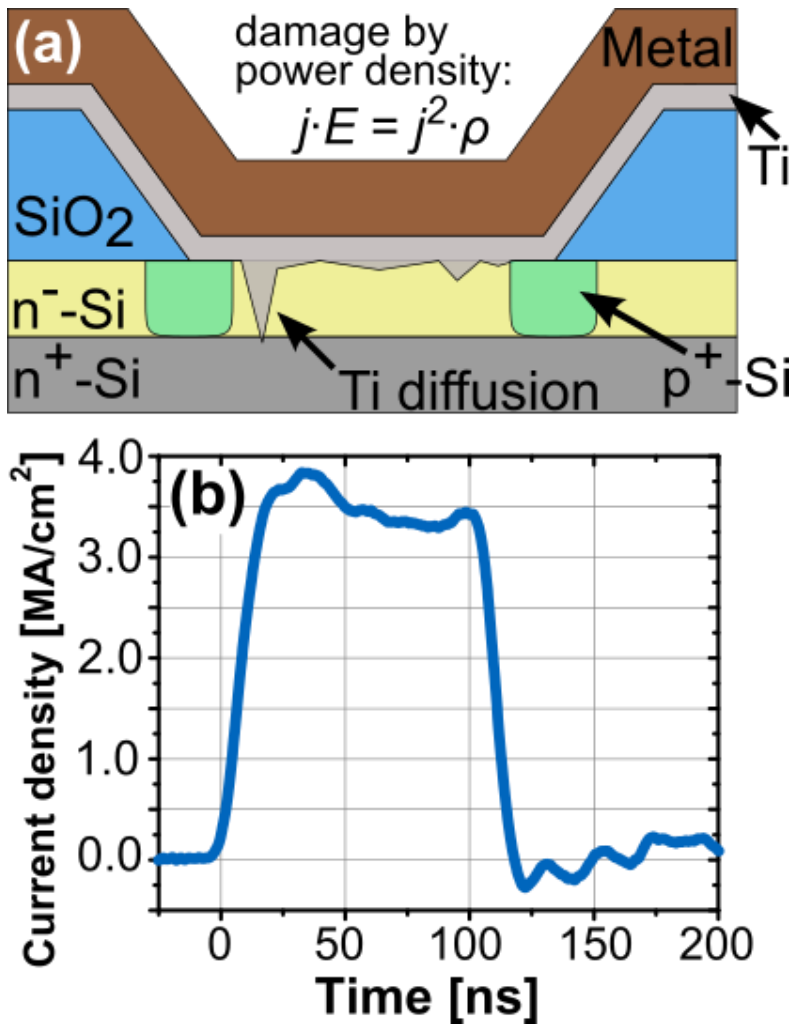

Fig. 2. (a) shows a schematic cross-section of a TiSi Schottky diode with guard ring structure where Ti diffuses into n-Si by the application of a high current pulse. Image (b) shows a typical waveform of a 100 ns current pulse with a rise/fall time of ~15 ns.

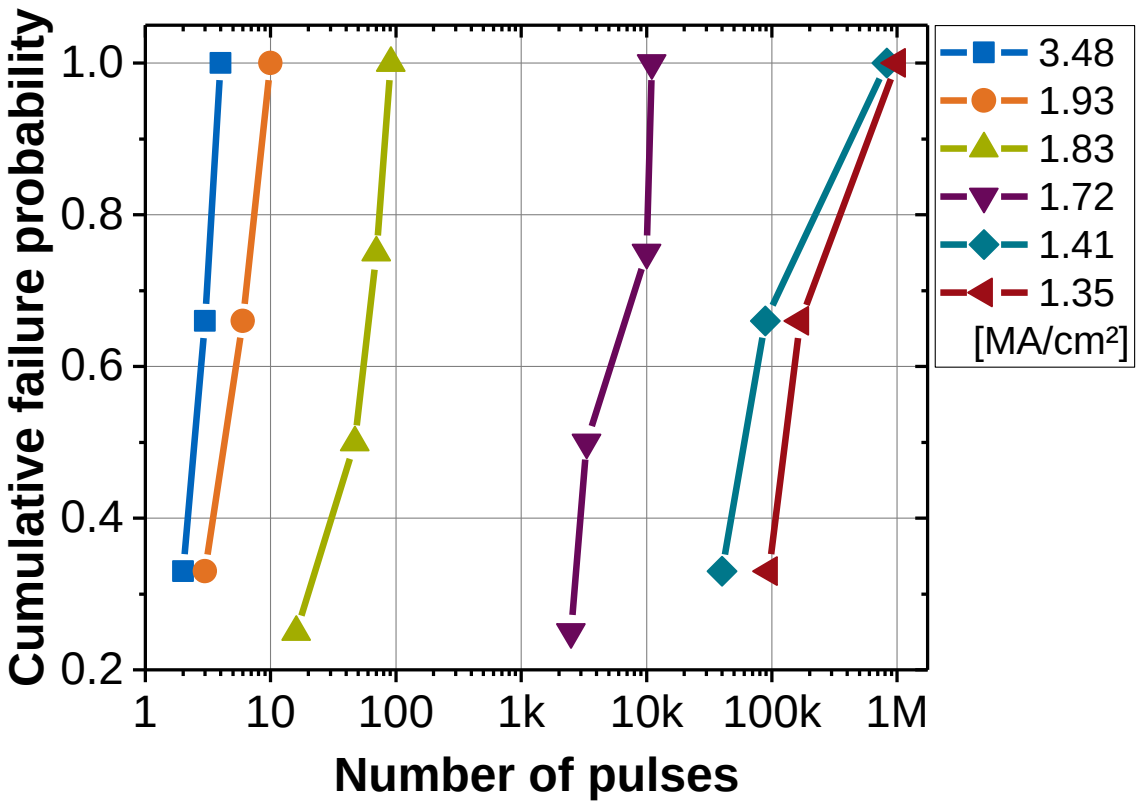

Fig. 3. Failure probability versus number of current pulses at different current densities for commercial BAT15 diodes with TiSi interface. At least three devices were stressed for a given pulse current density, a pulse width of 100 ns and a duty cycle < 0.0001. The failure of the diode was defined when the diode degrades to a reverse current >100 µA @ 1 V. The minimum current level is above the permissible pulse load of 1.21 MA/cm$^2$ specified by the manufacturer.

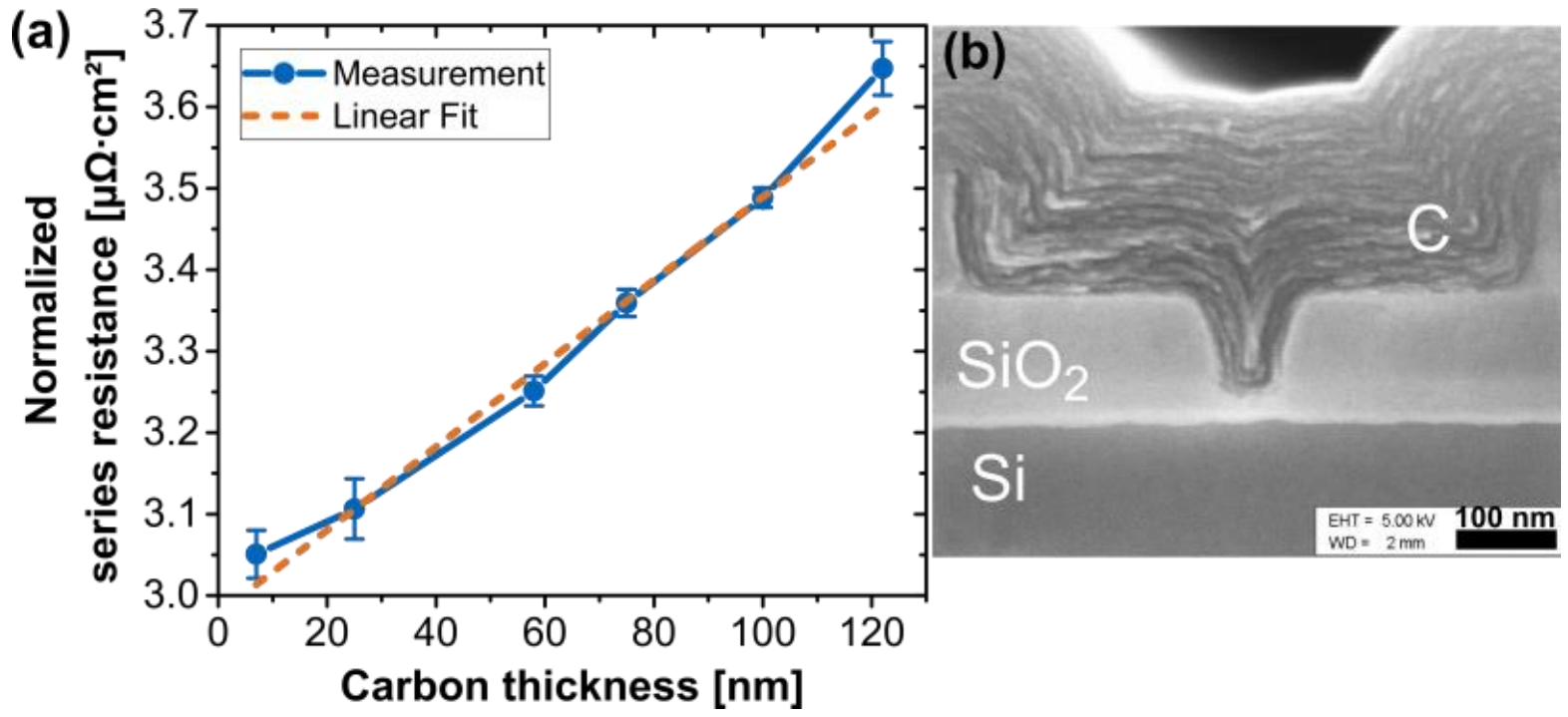

Fig. 4. The area-normalized series resistance of the C-Si Schottky diode as a function of the graphenic carbon thickness on the diode with the same metallization layers on top is shown in (a). The out-of-plane electrical resistivity is extracted to be 50 mΩ·cm. The laminar growth of graphenic carbon is revealed in the cross-section SEM in (b) of a cleaved sample, which displays the graphenic carbon layers. Due to the anisotropy, the resistivity in the perpendicular direction is 50x of the in plane value which is 1 mΩ·cm.

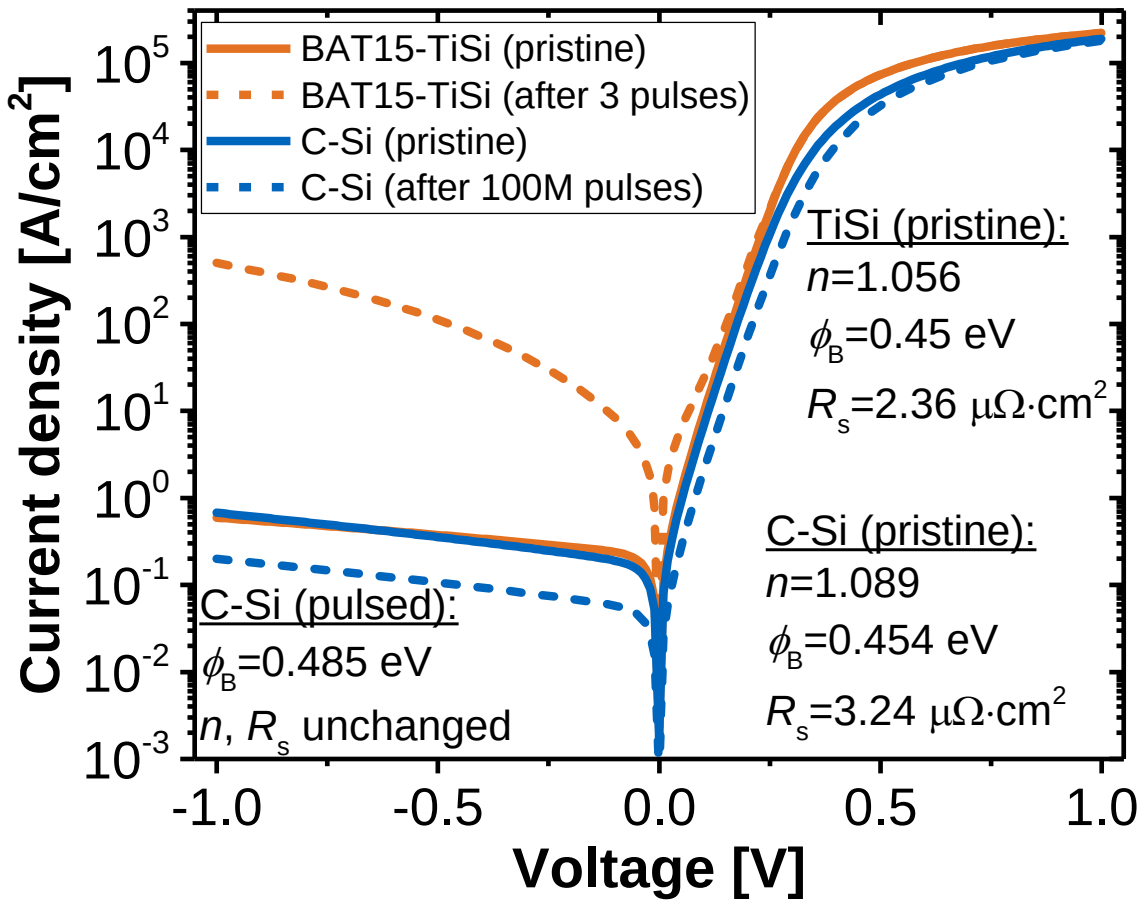

Fig. 5. Comparison of the dc characteristics of the BAT15 (TiSi) and the C-Si diode with a carbon thickness of 58 nm. Ideality factor $n$, Schottky barrier height $\Phi_B$, and area-normalized series resistance $R$s are displayed for the pristine BAT15 and C-Si diode. The barrier height is almost equal for both devices. After current pulses with a current density of 3.5 MA/cm$^2$, the TiSi shows a strong degradation in the reverse characteristics after 3 pulses while the C-Si diode has even a lower reverse current and is still fully functional after 100 million pulses.

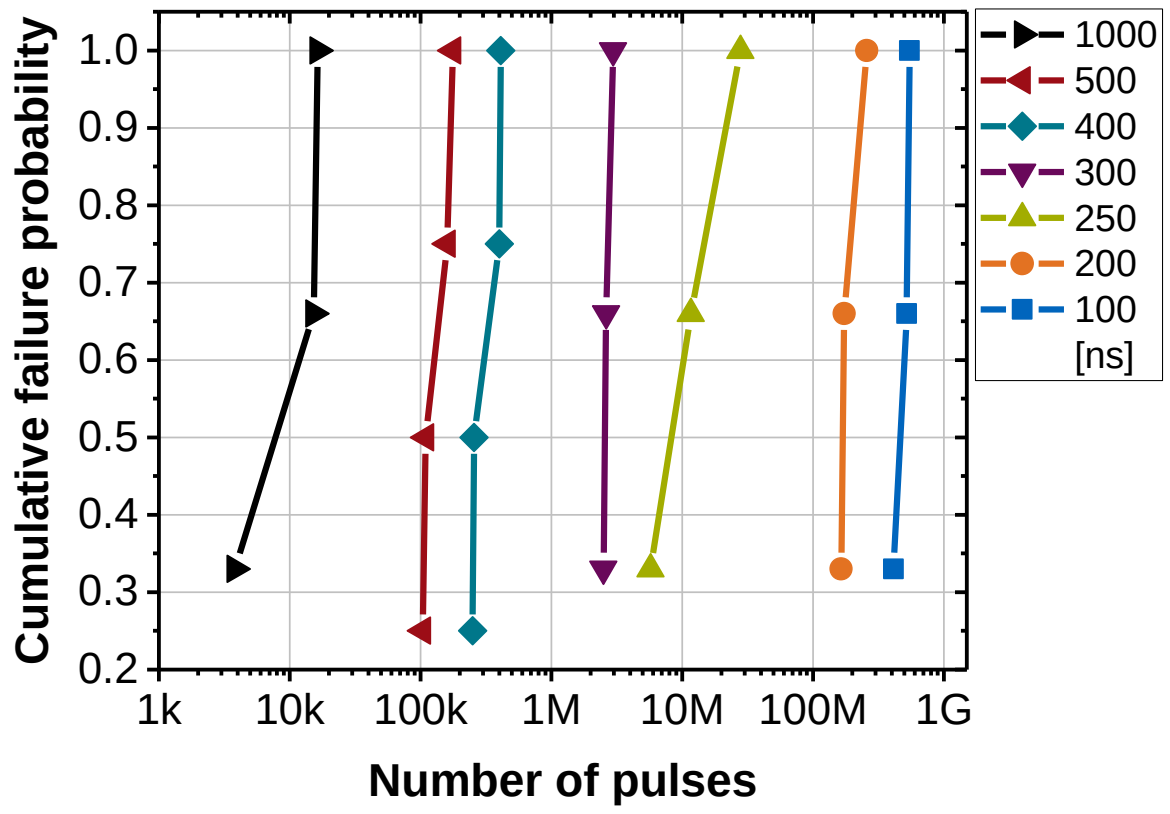

Fig. 6. Failure probability is plotted as a function of the number of pulses at different pulse lengths for the C-Si Schottky diodes. At least three devices were stressed with a given pulse current density of 3.5 MA/cm$^2$ and a duty cycle < 0.0005. The failure of the diode was defined when the diode degrades to a reverse current >100 µA @ 1 V.

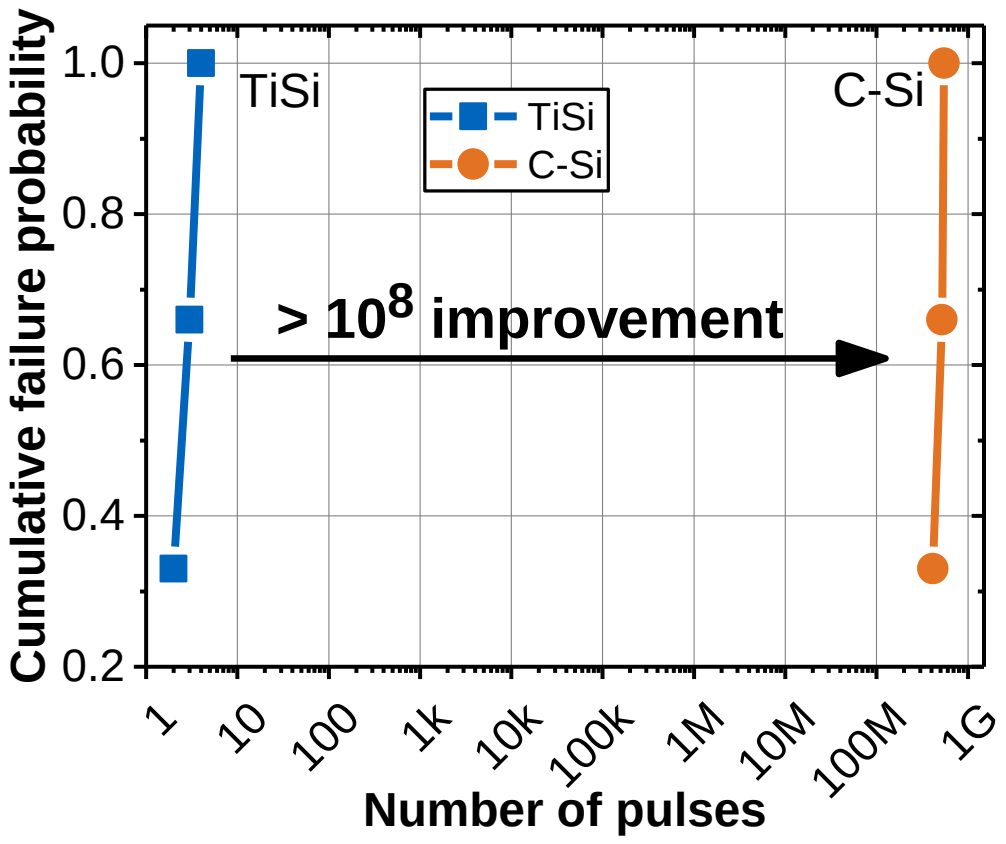

Fig. 7. Comparison of stress pulse endurance of TiSi and C-Si diodes for a current density of 3.5 MA/cm$^2$ and a pulse width of 100 ns. The majority of TiSi (BAT15) diodes is deteriorated after 3 pulses whereas the C-Si diodes can withstand **over $10^8$ more pulses** until a wear out is observed. This direct comparison underlines the importance of the C-Si contact scheme.

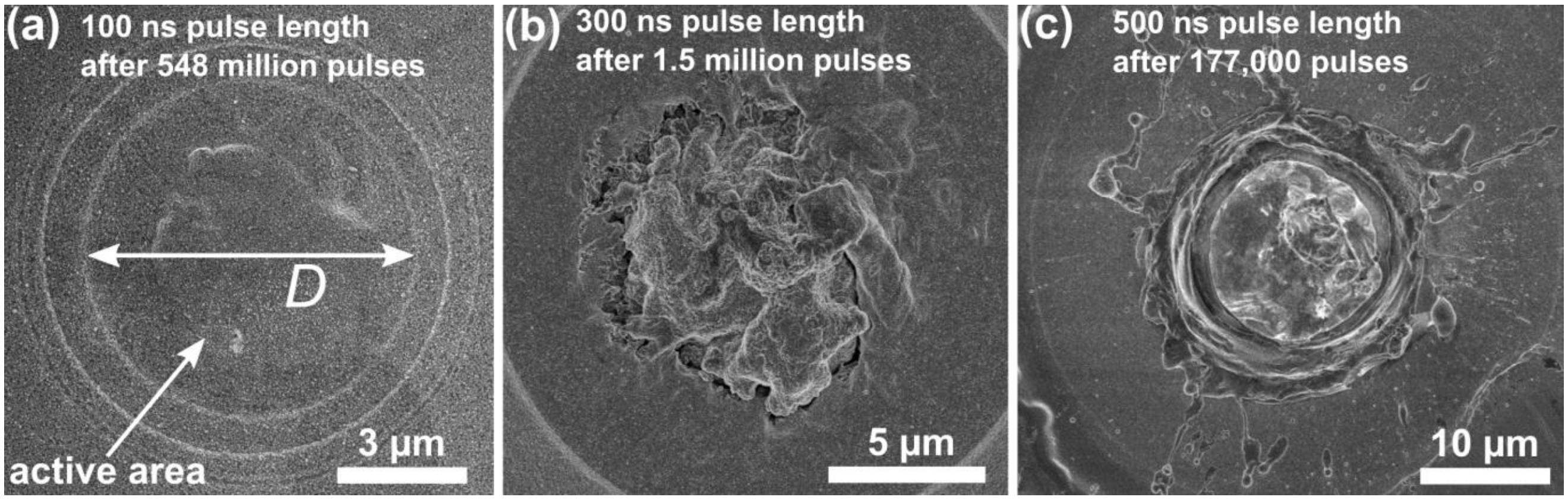

Fig. 8. SEM image of the top view of carbon-silicon Schottky diodes with copper as top metallization. All the devices where stressed with the same current level of 3.5 MA/cm$^2$ while only the pulse length was altered. The images were recorded after the diodes failed. (a) shows a device after 548 million pulses of a length of 100 ns where almost no damage in the top metallization is visible. The device in (b) demonstrates the damage of a 300 ns pulse after 1.5 million pulses where the metal even started to melt and to pile up. (c) illustrates a sample where the current-induced damage of a 500 ns pulse melted the metal so far that it spit on the periphery but the device could even withstand 177,000 pulses.

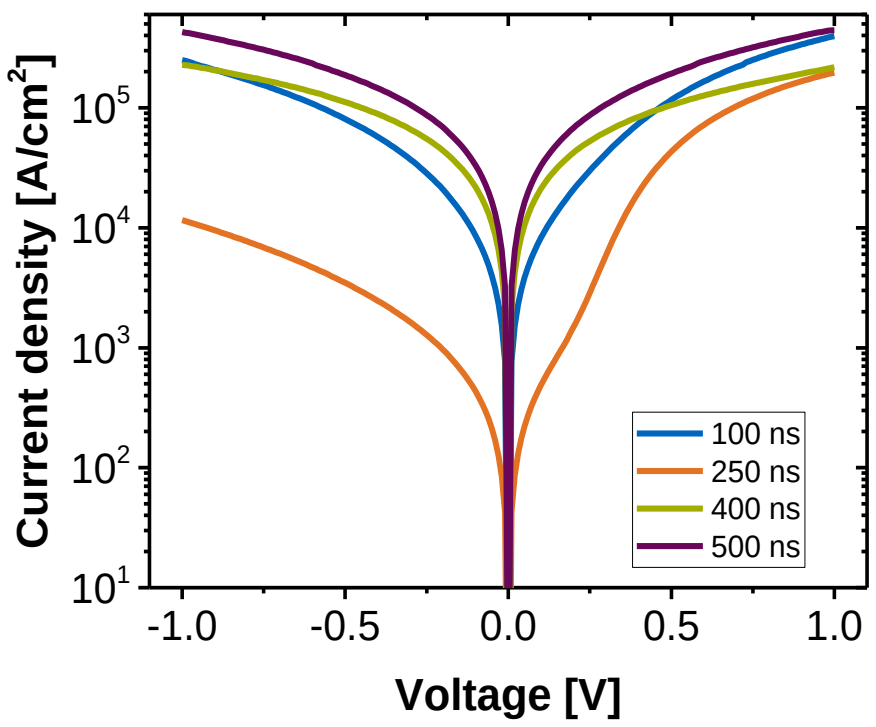

Fig. 9. J(V)-characteristics of C-Si diodes after they reached the threshold for a failed diode during the test in Fig. 6. The diodes show an increased $J_0$ and consequently a reduced Schottky barrier height.

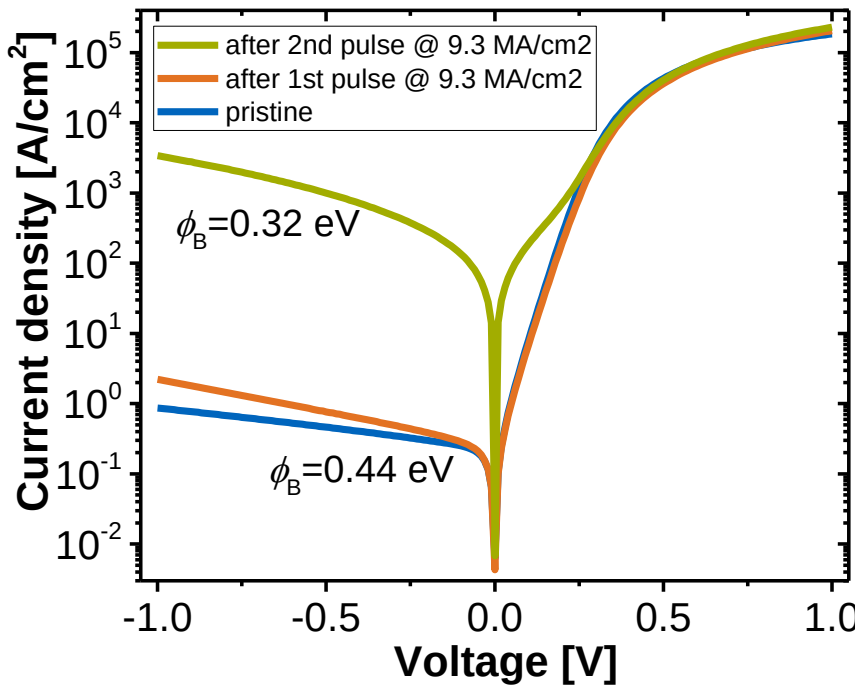

Fig. 10. J(V)-characteristics of a C-Si diode with a carbon thickness of 58 nm. After one pulse with a current density of 9.3 MA/cm$^2$ the diode is fully functional and in its reverse behavior still close to the pristine curve. Only the second pulse is destructive.

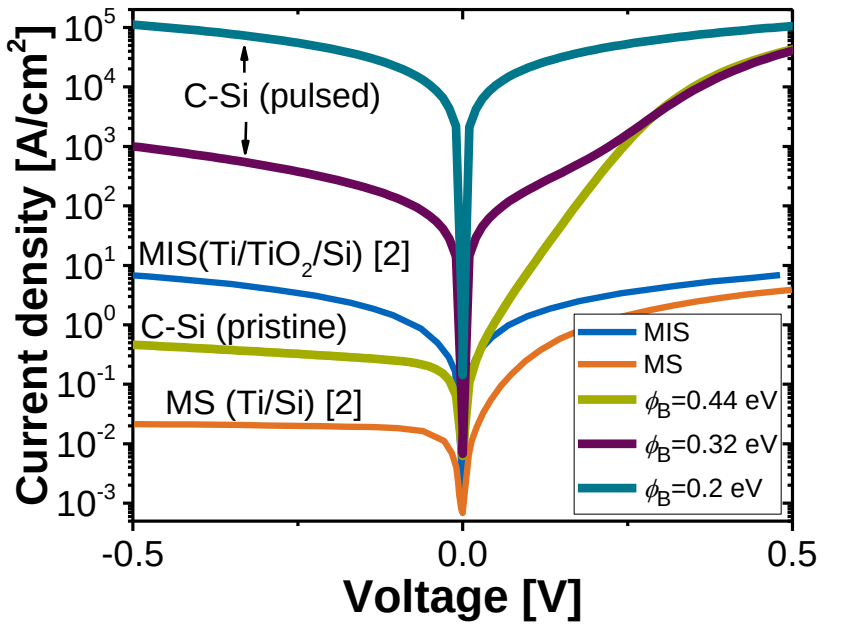

Fig. 11. Comparison of current density plots J(V) of literature values of MS (Ti/Si) and MIS (Ti/TiO$_2$/Si) with C-Si values. The contact resistance $\rho_c$ is inversely proportional to $J_0$ which is even higher for pristine C-Si compared to the MS diode from [2]. The pulse-annealed C-Si diodes are even much better.

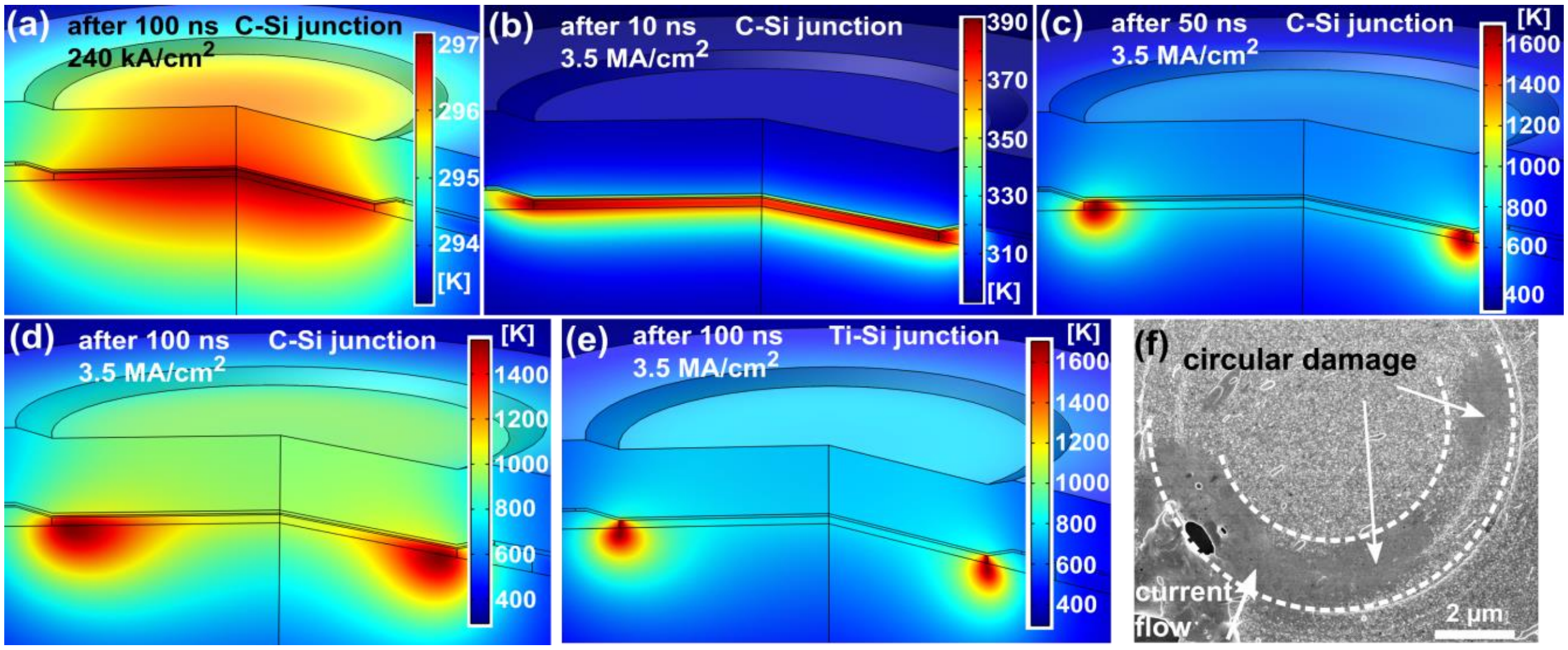

Fig. 12. Time-dependent electro-thermal simulations of the used Schottky-contacts. Image (a) shows the temperature in the device operated at normal conditions (240 kA/cm$^2$). In (b)-(d), the temperature distribution in C-Si diode is shown when stressed with a 3.5 MA/cm² current pulse. A circumferential hot-spot is created in the epi-layer during the pulse, but the heat spreads out which cools down the hot spot under the melting temperature. This is due to the unusual temperature dependent electrical resistivity which has a minimum at around 900 K and increases again at higher temperature. In (e) titanium (Ti) is used as interface metal. Here, the hot spot doesn't disappear during the pulse which can lead to a more severe damage to the device. The SEM image in (f) shows a C-Si diode with thinner metallization after several 3.5 MA/cm$^2$ pulses. The direct current path from the probe tip to the diode is indicated by the arrow. A recrystallization of the top metallization at the transition region between guard ring and active epi region occurs. This indicates that the main damage is indeed created in a circular fashion around the diode active area, like identified in the FEM simulations.